\documentclass[prl,reprint,superscriptaddress,nobalancelastpage,showpacs]{revtex4-1}
\usepackage[normalem]{ulem}
\usepackage{pdfpages}
\usepackage{cancel}
\usepackage{graphicx}
\usepackage{amssymb}
\usepackage{natbib}
\usepackage{dcolumn}
\usepackage{bm}
\usepackage{subfigure}
\usepackage{graphicx,psfrag,xspace}
\usepackage{color}
\usepackage{amssymb,amsfonts,amsmath}
\usepackage{setspace}
\usepackage{datatool}
\def\bra#1{\mathinner{\langle{#1}|}} 
\def\ket#1{\mathinner{|{#1}\rangle}} 

\def\omw{\omega_{\text{\tiny MW}}}
\def\stat{\infty}
\newcommand{\titleinfo}{Dynamic nuclear polarization and the paradox of Quantum Thermalization}

\def\omw{\omega_{\text{MW}}}

\graphicspath{{figure/}}

\begin{document}

\author{Andrea De Luca  and Alberto Rosso}
\affiliation{Laboratoire de Physique Th\' eorique et Mod\` eles Statistiques (UMR CNRS 8626), Universit\' e Paris-Sud, Orsay, France}

\title{\titleinfo}
 
\begin{abstract} 
Dynamic Nuclear Polarization (DNP) is to date the most  effective technique to increase the 
nuclear polarization opening disruptive perspectives for medical applications. 
In a DNP setting, the interacting spin system is quasi-isolated and brought out-of-equilibrium
by microwave irradiation. 
Here we show that the resulting stationary state strongly depends on the 
ergodicity properties of the spin	 many-body eigenstates. In particular
the dipolar interactions compete with the disorder induced by local magnetic fields 
resulting in two distinct dynamical phases: while for weak interaction, 
only a small enhancement of polarization is observed,
for strong interactions the spins collectively equilibrate to an extremely low effective temperature 
that boosts DNP efficiency. 
We argue that these 
two phases are intimately related to the problem of thermalization 
in closed quantum systems where a many-body localization transition 
can	 occur varying the strength of the interactions.
\end{abstract}


\maketitle

\begin{figure}[t]
\centering
\includegraphics[width=0.95\columnwidth]{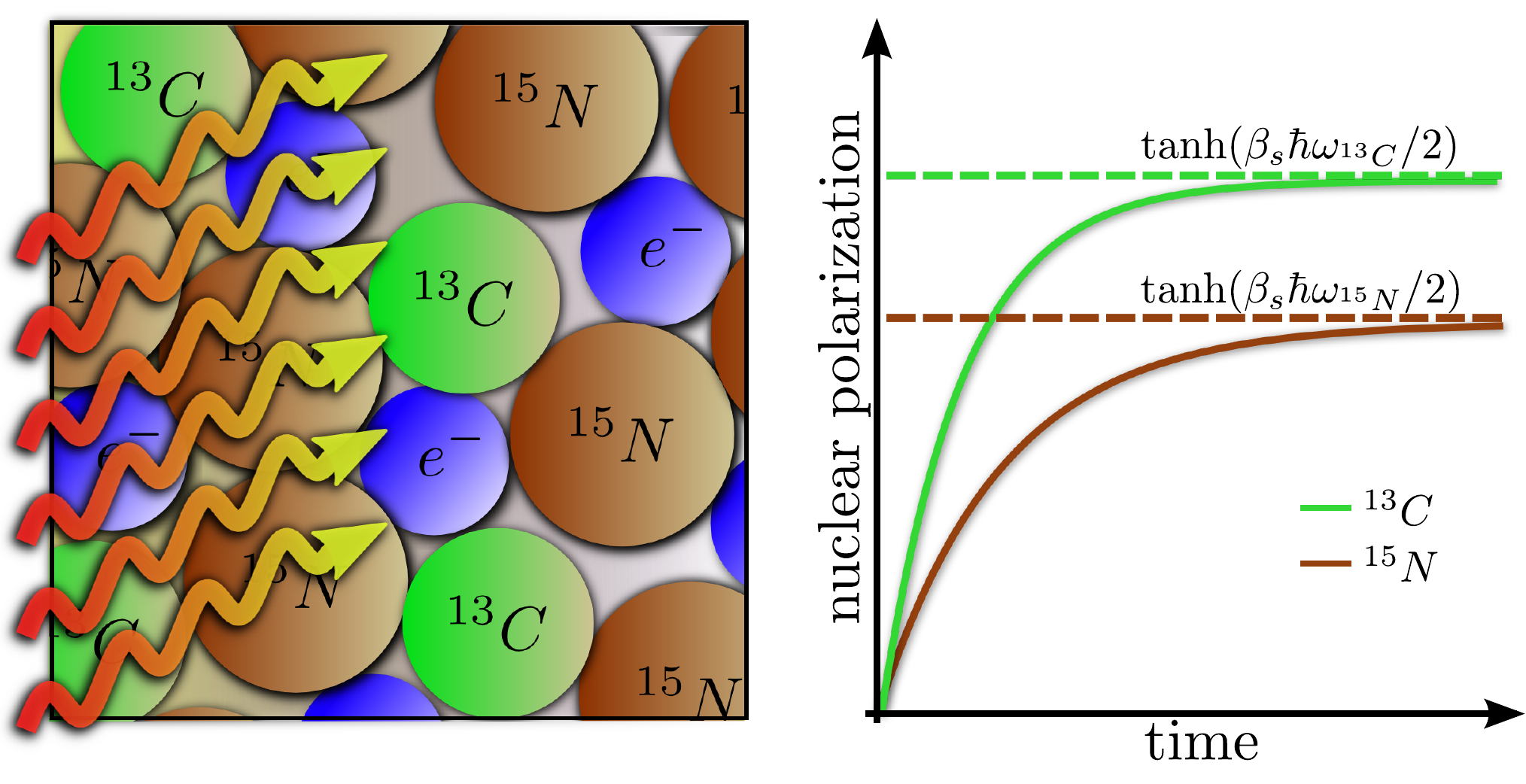}
\caption{
\label{DNPsketch}
Color online. A solid material containing nuclear spins (e.g. $^{13}C$, $^{15}N$)  and doped with 
molecules with unpaired electrons (left). At $1.2$ Kelvin and  $3.35$ Tesla the equilibrium polarization 
of the electron spins is very high, ($94\%$), while nuclear spins are very little polarized, 
less than $1\%$. Under microwave irradiation the spin system evolves towards a new steady state
characterized by a single spin temperature $\beta_s^{-1} \sim 1 $ mK (right). 
In this work, we analyze exclusively the electron spins and show that 
an out-of-equilibrium spin temperature results from the interplay of disorder and interaction.
}

\end{figure}

\paragraph{Introduction. ---}
 Formulating statistical mechanics for isolated many-body system
requires the tacit assumption of ergodicity. Only recently, however,
the eigenstate thermalization hypothesis (ETH)  \cite{rigol2008thermalization, biroli2010effect}
promoted this concept to a testable condition at the quantum level, allowing the identification of
situations where ergodicity might even be broken once disorder combines with quantum interference   \cite{basko2006metal, pal2010many}.
It is natural to ask, then, whether ETH
influences also the stationary regimes where energy is constantly injected 
and dissipated, leading again to an emergent simple description.
Dynamic Nuclear Polarization (DNP), the most effective technique to increase the 
nuclear polarization, is a paradigmatic out-of-equilibrium protocol to test these ideas.  
In a DNP procedure \cite{abragam1978principles}, the
compound is doped with radicals (i.e., molecules with unpaired electrons), exposed to a strong magnetic field
at low temperature, $\beta^{-1}$,  and then irradiated with microwaves (see Fig.~\ref{DNPsketch} for details). At thermal equilibrium, the unpaired electrons are much more polarized than nuclear spins
because the electron Zeeman gap is orders of magnitude larger than the nuclear one. When the microwaves are on,
at a frequency close to the electron Zeeman gap, the spin system of interacting electrons and nuclei organizes
in an out-of-equilibrium steady state with a huge nuclear polarization.
The hyperpolarized sample can then be dissolved at room temperature  \cite{ardenkjaer2003increase},
injected in patients, and  used as metabolic tracer  \cite{golman2006real}.
However, our understanding of the  physical mechanisms that trigger hyperpolarization is still poor.
A striking experimental evidence is the thermal mixing of the ensemble of 
different nuclear spins  ($^{13}C$, $^{15}N$, $^{89}Y$, $\ldots$) \cite{lumata2011dnp, kurdzesau2008dynamic}: 
their enhanced polarizations are well described by an equilibrium-like polarization,  $P_n = \tanh(\beta_s \hbar \omega_n /2)$ (see Fig.~\ref{DNPsketch} right). 
While the Zeeman gap $\omega_n$ depends on the nuclear species,  the spin temperature $\beta_s^{-1}$ is a unique parameter, 
possibly one thousand times smaller  than $\beta^{-1}$, the  one of the bath  \cite{provotorov1962magnetic}.

But how can a quantum system appear thermal and colder when irradiated by microwaves?
In which way the spin temperature can be controlled acting on the experimental parameters? 

In this paper we show that the spin temperature concept  is directly connected 
to quantum ergodicity and ETH. While for classical physics, thermalization has its origin
at the onset of chaotic dynamics, quantum ergodicity requires the ETH, a thermal behavior at the level of single eigenfunctions  \cite{rigol2008thermalization, polkovnikov2011colloquium}.
The  realm of ETH is normally restricted to quench protocols in cold atoms experiments, 
where any exchange of energy with the enviroment is under control.
Our work shows that ETH may impose a thermal behavior to the stationary state of open quantum systems, giving  a practical and experimental relevance to the fundamental problem of Quantum Thermalization 
\cite{basko2006metal, pal2010many}.


\begin{figure}
\centering
\includegraphics[width=0.85\columnwidth]{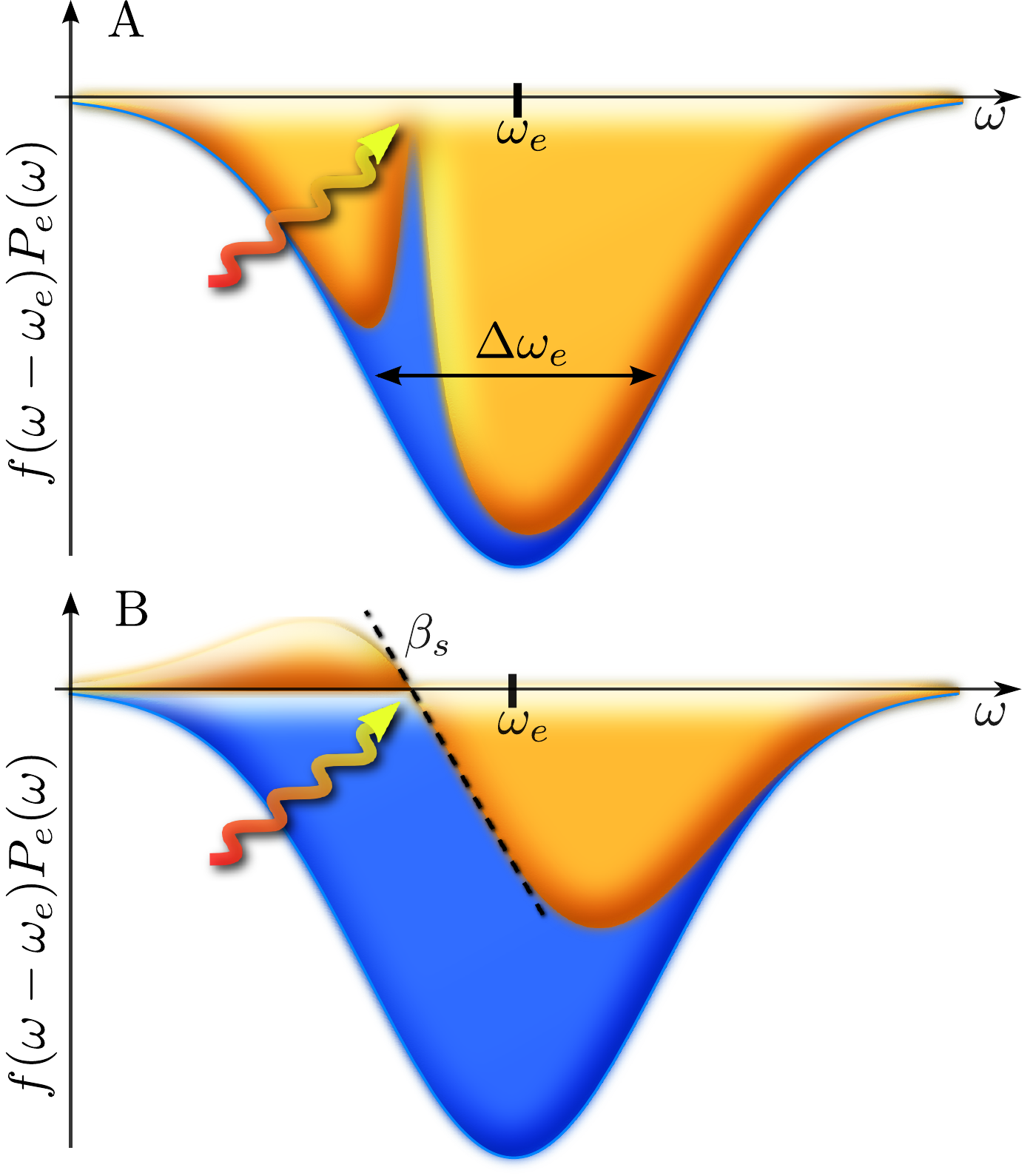}
\caption{\label{EPRsketch}
\textbf{EPR spectrum} Equilibrium (blue) versus MW irradiated (yellow) profile. 
Under irradiation two possible profiles are expected:
(A) the hole burning shape, characteristic of the non interacting case  (B) the hyperbolic tangent shape 
characterized by a very low effective  temperature, $\beta_s$, that cools down the nuclear spins.}
\end{figure}

\paragraph{The microscopic model. ---}
The traditional description of DNP in the thermal mixing  regime relies on the phenomenological assumption
that the electron spins cool down once irradiated and act as a reservoir for all nuclear species.
Here, we focus only on the electron spins and on the origin of the spin temperature 
in their stationary state. In the electron spin Hamintonian, 
the presence of g-factor anisotropy induces a spread of the electron Zeeman gap:
 \begin{equation}
 \label{hamiltonian}
 \hat H =  \hbar \sum_{i =1}^ N \left( \omega_e +\Delta_i \right) \hat S_z^i + {\hat H}_{\text{dip}} \; .
 \end{equation}
where $N$ is the number of electrons and $\omega_e$ is the external magnetic field.
The random fields $\Delta_i$ are quenched at low temperature and distributed according to the density $f(\Delta)$, with mean $\overline{\Delta_i} = 0$ and variance $\overline{\Delta_i^2}=\Delta\omega_e^2$. 
The term ${\hat H}_{\text{dip}} $ contains the interactions between spins due to the dipolar coupling. 
Experiments can access the product $f(\Delta) P_e(\omega)$, dubbed EPR spectrum, 
where $P_e(\omega)$ is the polarization of an electron   with Zeeman gap $\omega=\omega_e + \Delta_i$. 
At equilibrium  with the environment (sketched in blue in Fig.~\ref{EPRsketch}), $P_e(\omega) \simeq -1$. 
The microwave irradiations at frequency $\omw$ and intensity $\omega_1$ takes the form 
$\hat H_{\text{\tiny MW}} = 2 \omega_1 \hat \sum_{i} S_x \cos(\omw t)$, with $\hat S_x$ total spin operator along the $x$ component.
In absence of dipolar interactions,
the Bloch equations predict that the electrons  with a resonating Zeeman gap are saturated, 
$P_e\left(\omega \sim \omega_{\text{MW}}\right) \sim 0$,  while the others
remain highly polarized. This corresponds to the {\em hole burning shape} of EPR spectrum, showed in Fig.~\ref{EPRsketch}A.
On the contrary, according to the thermal mixing picture, the presence of dipolar interactions induces a collective 
reorganization of the electron polarization profile $P_e(\omega)$, that shows an equilibrium-like shape even under microwave irradiation
\begin{equation}
\label{ansatz}
P_e\left(\omega\right) = - \tanh\left[\frac{\hbar \beta_{\text{s}}}{2} 
(\omega-\omega_0)\right]
 \end{equation}
with $\omega_0 \simeq \omw$. 
The ansatz of Eq. \eqref{ansatz} lacks a microscopic derivation. Moreover, recent
ab-initio models \cite{hovav2015electron} have only observed an hole burning shape, with 
a weak polarization enhancement triggered by local hybridizations
\cite{hovav2010theoretical,karabanov2012quantum, hovav2012theoretical}. 
 Here, we  take
 \begin{equation}
 \label{hamiltonian2}
   {\hat H}_{\text{dip}} = \sum_{i<j} A_{ij} \left( \hat S^i_+\hat S^j_- + c.c.    \right) \; .
 \end{equation}
  where the $A_{i,j}$ are the dipolar couplings. Because of the glassiness of DNP samples, 
the distance between electrons is random and, thus, for simplicity  
the coupling $A_{ij}$ are  taken, within a mean field approximation,  as gaussian random variables 
with zero mean and variance $U^2/N$.  We are interested in the strongly correlated regime where
disorder and interaction compete, i.e. $U \simeq \Delta \omega_e$.
Our conclusions should not depend on the specific model 
and, here,  we  choose a uniform distribution of local magnetic fields
by taking equally spaced $\Delta_i = \Delta \omega_e \left( \frac{2i -  N-1}{2N} \right)$,
 with randomness only affecting the dipolar couplings.

\begin{figure}[t]
\includegraphics[width=1.0\columnwidth]{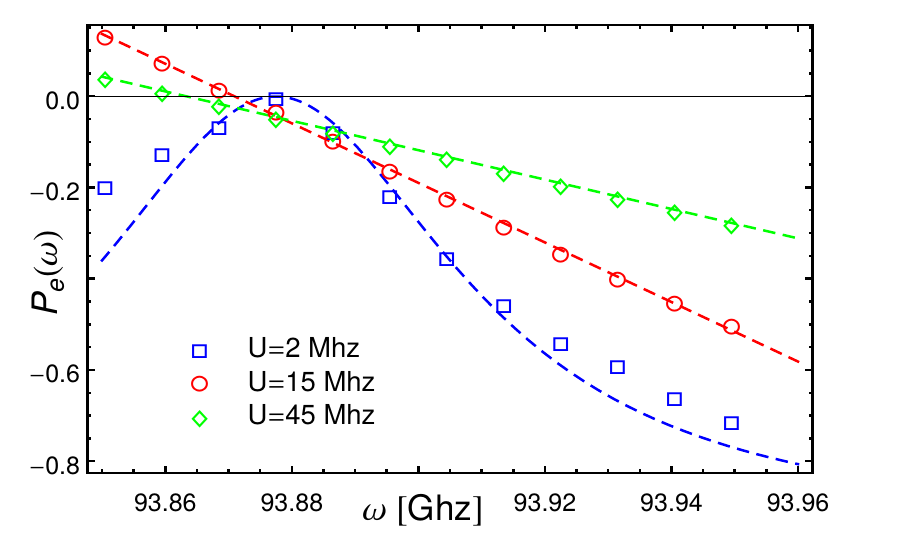}
\caption{\label{polarization}
Electron polarization under MW irradiation with $\omw=93.8775$ Ghz, 
for the model of Eq.\eqref{hamiltonian} with $N=12$ spins. For strong dipolar 
interactions (circles and diamonds)
the spin temperature is well defined. The fit according to Eq. 
\eqref{ansatz} (red and green lines) 
gives $\beta_s^{-1}= 3.7$ mK for $U=15$ Mhz and $\beta_s^{-1}=7.4$ mK for $U=45$ 
Mhz. 
For weak interactions $U= 2$ Mhz (square) a simple broadening of the hole 
burning 
non interacting profile (dashed blue line) given in Eq.~(8) of \cite{Note2}.
}
\end{figure}

\paragraph{The master equation. ---} 
A key experimental observation \cite{johannesson2009dynamic, filibian2014role}
is that the spin system is quasi-isolated with a dephasing time, 
$T_2$, very short compared to the time-scales of microwave dynamics 
and to the relaxation time, $T_1$, with the thermal bath
\footnote{for the electron spins of trytil radicals  at 1.2 Kelvin and 3.35 Tesla we have $T_2 \simeq 10^{-6}$ sec. versus $T_1 \simeq 1$ sec.}.  
Therefore, any initial density matrix,  $\rho$ is quickly reduced by dephasing 
to a diagonal form in the basis of eigenstates of $\hat H$.
In practice \footnote{See Supplemental Material, which includes Refs. [36--39]}, the Lindblad equation 
$\dot\rho = \mathcal{L}\rho$
used to describe the dynamics of the open system reduces to a master equation 
for the time evolution of the $2^N$ occupation probabilities, $\rho_{nn}$
with rates $W_{n \to n'} =   h(\Delta \epsilon_{n,n'}) W^{\text{bath}}_{n, n'} +  
W^{\text{MW}}_{n, n'}$, where
\begin{subequations}
\label{matrixtransition}
\begin{align}
\label{matrixtransitionBATH}
&W^{\text{bath}}_{n, n'} =  \frac{2}{T_1}\sum_{j=1}^N \sum_{\alpha=x,y,z} |\bra{n}\hat S_{\alpha}^{j}\ket{n'}|^2\;,\\
\label{matrixtransitionMW}
&W^{\text{MW}}_{n, n'} = \frac{4 \omega_1^2 T_2 |\bra{n}\hat S_{x}\ket{n'}|^2}{1+ T_2^2 (|\Delta\epsilon_{nn'}| - \omw )^2}\;.
\end{align}
\end{subequations}
Here the index $n$ label eigenstate of energy $\epsilon_n$
with $\Delta\epsilon_{nn'} = \epsilon_n - \epsilon_{n'}$.
 The function $h(x)=e^{\beta x}/(1+ e^{\beta x})$ 
assures the convergence to Gibbs equilibrium when the 
microwaves are off and the rate 
$W^{\text{bath}}_{n, n'}$  in \eqref{matrixtransitionBATH} comes from single spin flip transitions 
on a time scale $T_{1}$.
 Eq.~\eqref{matrixtransitionMW} describes transitions induced by the microwave field. 
In Fig.~\ref{polarization} we present the stationary value  of the polarization  
$P_e(\omega = \omega_e + \Delta_i) \equiv 2\operatorname{Tr}[S_{z}^i \rho_{\stat}]$, 
 computed from the stationary occupation probabilities which solve $\mathcal{L}\rho_{\stat} = 0$. 
Note that this requires the full diagonalization of $\hat H$, strongly constraining the possible system sizes.

 \begin{figure*}[ht]
\begin{minipage}{0.3\textwidth}
\centering
\includegraphics[height=0.874\textwidth]{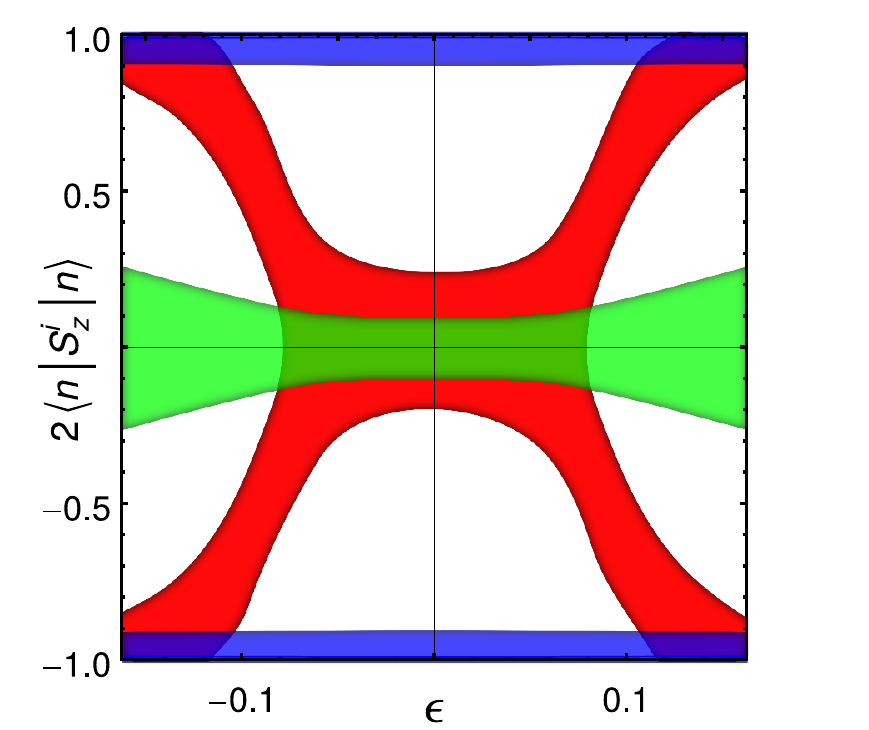}
\end{minipage}
\begin{minipage}{0.3\textwidth}
\centering
\includegraphics[height=0.874\textwidth]{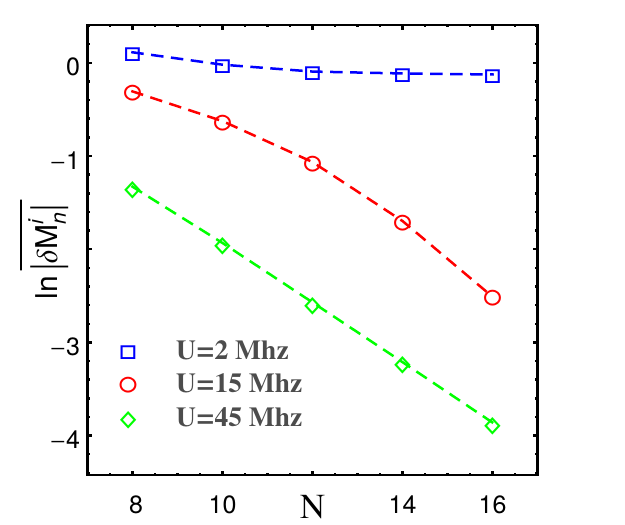}
\end{minipage}
\begin{minipage}{0.38\textwidth}
\centering
\includegraphics[height=0.674\textwidth]{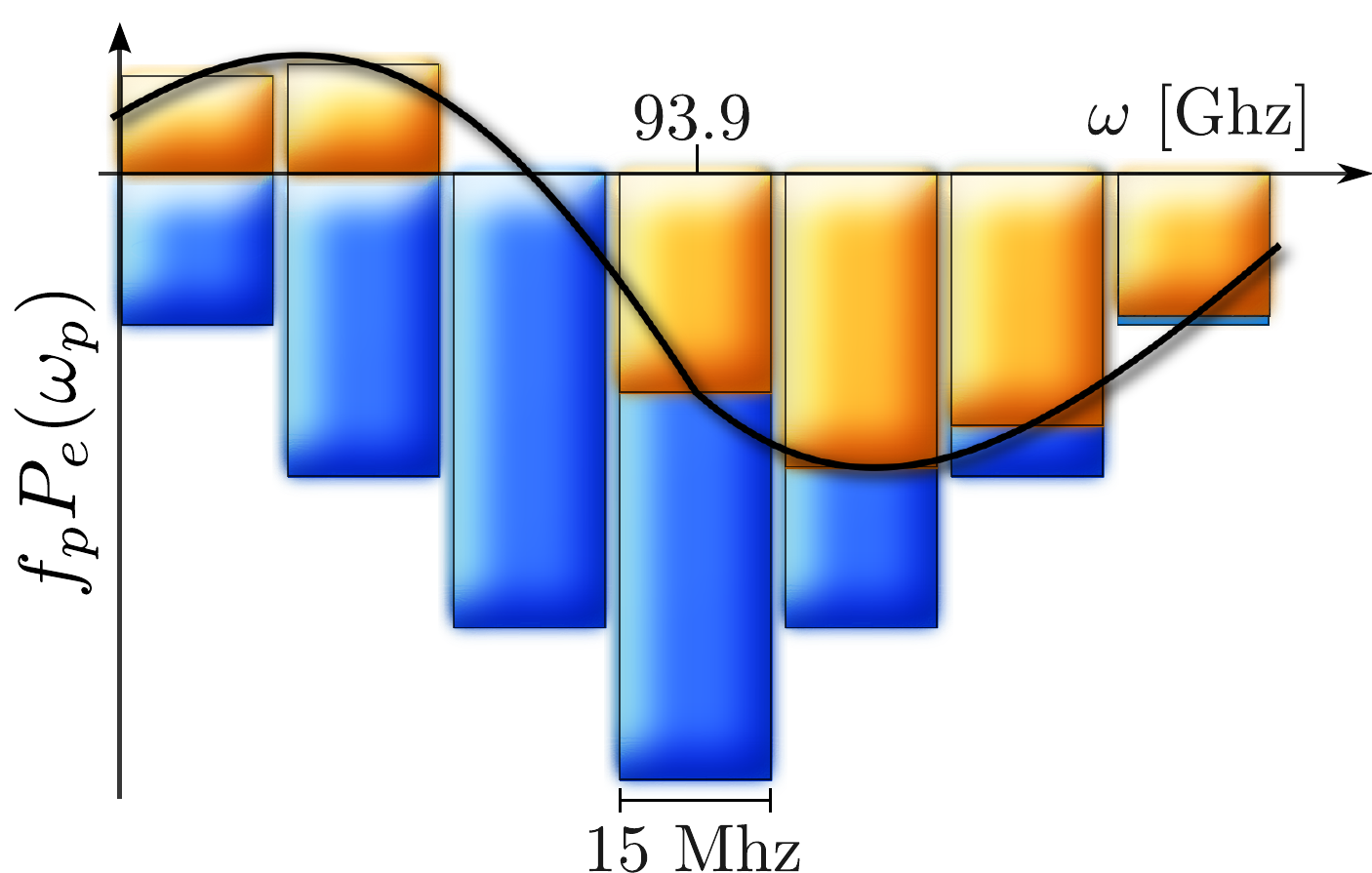}
\end{minipage}
\caption{\label{ETHsketch} Left: 
Density plot of the distribution of diagonal elements $\bra{n} \hat S_z^i \ket{n}$ 
in the sector of vanishing total polarization. Colored regions
represents, for each energy window $(\epsilon, \epsilon + \Delta \epsilon)$, the smallest area containing half probability ($U= 2$ Mhz in blue, $U= 15$ Mhz in red, $U= 45$ Mhz in green). 
Middle: Logarithm of the variation of the local polarization between pairs of adjacent eigenstates of Eq.\eqref{hamiltonian} 
versus $N$. In the ergodic phase, this indicator vanishes exponentially in $N$. In the localized phase, 
it saturates to a finite value. Right: EPR spectrum for the toy model of Eq.~(\ref{hamiltoniantoy}) with $7$ packets. 
The equilibrium profile is in blue. The yellow histograms show the stationary profile under MW irradiation with $\omw=93.8775$ Ghz and 
$N=64$ spins. The solid line is obtained from the Ansatz of Eq. \eqref{ansatz} imposing the condition \eqref{conservation}. }
\end{figure*}

Two possible behaviors are observed: 
For weak interactions, the hole 
burning shape, already observed in \cite{hovav2010theoretical,karabanov2012quantum, hovav2012theoretical}, is recovered.
Instead, in presence of strong dipolar interactions, we show that 
all electrons rearrange according to the spin temperature profile of \eqref{ansatz}. 
Remarkably, in the wing closer to $\omega_{\text{MW}}$, electron polarization can  even invert 
its sign. 

The origin of these two dynamical regimes can be understood in relation with
the quantum thermalization of the electron spins. 
In general, in closed quantum systems, 
an arbitrary initial state converges to
a time-independent density matrix because of
dephasing: in the basis of eigenstates, off-diagonal elements are suppressed
while the occupation probabilities on the diagonal remain constant.
But then, how could thermodynamics emerge if the initial occupancies are conserved?
In  Fig.~\ref{ETHsketch} (left) we report the most probable value of the 
local polarization for the eigenstates at a given energy. 
In presence of weak interaction ($U=2$ Mhz)
 the polarization fluctuates  between the extremal values $\pm 1$ showing that 
the exact eigenstates  are almost factorized on local spins.
 When the interaction is increased, eigenstates are strongly entangled
 and the local polarization is close to zero, its microcanonical average. 
  As predicted by
ETH,
 each eigenstate is independently thermal and so the paradox of Quantum Thermalization 
is solved, as the memory of the initial condition 
fades out while entanglement grows through dephasing \cite{amico2008entanglement}. 
Instead, in the weakly interacting regime, for initial states close enough 
to exact eigenstates,
a finite fraction of the polarization is doomed to survive \cite{buccheri2011structure, de2013ergodicity, ros2015integrals}.

In absence of disorder, the spin-temperature is well-defined but very high \cite{hovav2013theoretical};
varying the ratio $U/\Delta \omega_e$, the spin-temperature decreases up to a point where
the system approaches the many-body localization (MBL), a dynamical transition between an ETH and a non-ergodic phase 
\cite{oganesyan2007localization, basko2006metal, pal2010many}, surviving even in presence of microwaves \cite{ponte2015periodically}.
In Fig.~\ref{ETHsketch} (middle) we present a standard indicator for the 
transition: the variation of the local polarization 
between pairs of adjacent eigenstates versus the size of the system 
\cite{pal2010many}. 
In the ETH phase, this quantity converges exponentially to zero indicating that 
all the fluctuations 
are more and more suppressed. On the contrary, in the localized phase, it 
saturates to a finite value, 
since fluctuations remain present even in the thermodynamic limit.

Our results indicate that 
 whenever the interaction with the environment
is weak but not negligible, the dynamics reduces to quantum jumps 
between exact eigenstates of the electron system. Then, if ETH holds, the stationary state 
necessarily looks thermal, with few global parameters (e.g. the spin-temperature) 
fixed relaxation and microwave irradiation.
Instead, in the localized phase, only a weak DNP enhancement, triggered by 
few-body processes, can be observed.

\paragraph{Spin temperature behavior. ---} 
It is important now to estimate the value of the spin temperature in the ETH phase. We first study a simplified model where the electrons in the EPR spectrum are assumed to be grouped into well separated 
macroscopic packets:
 \begin{equation}
 \label{hamiltoniantoy}
 \hat H_{\text{toy}} =  \hbar \sum_{p} \left( \omega_e +\Delta_p \right) 
\sum_{k=1}^{N_p} \hat S_z^k  +\eta H_{\text{int}}
 \end{equation}
 where $\sum_p N_p \Delta_p =0$ and $N_p$ is the number of  electrons in the 
packet $p$. For $\eta=0$ the spectrum of the Hamiltonian is composed by sectors 
of defined total magnetization  and energy.  The interactions are encoded in 
$H_{\text{int}}$, which is chosen as a Gaussian random matrix inside each 
sector. 
 When $\eta$ is small -- but still prevailing over the coupling with the bath --
 $H_{\text{int}}$ lifts the degeneracies in each sector
 selecting an ergodic basis in which the long-time density matrix is diagonal 
\cite{brandino2012quench}. 
 This model allows avoiding the numerical diagonalization, since
statistical properties of the eigenstates are known.
Moreover, the rates $W_{n\to n'}$ in Eq.~\eqref{matrixtransition} depend on 
$n,n'$ only via the matrix elements of local spin operators, 
$|\bra{n} \hat S_x^i \ket{n'} |^2$. Being the eigenvectors perfectly ergodic
in each sector, this quantity is actually determined by  the pair of sectors containing respectively $n,n'$, 
with weak statistical fluctuations \cite{Note2}. 
These simplifications largely reduce the exponential difficulty of the problem.
In Fig \ref{ETHsketch} (right) we show that the stationary EPR spectrum for 
$N=64$ spins perfectly agrees with the thermal Ansatz in Eq.~\eqref{ansatz}.
Moreover, $\beta_s$ and $\omega_0$ in \eqref{ansatz} can be fixed
imposing that the energy and total magnetization become stationary for large times,
which for the toy model leads to \cite{Note2}
\begin{subequations}
\label{conservation}
\begin{align}
 &2 T_2 \omega_1^2 P_e(\omw) 
 + \sum_p N_p  \frac{P_e(\omega_p) - P_0}{2T_1} 
 =0 \;, \\
 &2 T_2 \omega_1^2 \Delta_0 P_e(\omw) 
 + \sum_p N_p \Delta_p \frac{P_e(\omega_p) - P_0}{2T_1} 
 =0 \;.
\end{align}
\end{subequations}
where $\omega_p = \omega_e + \Delta_p$, $P_0 = -\tanh(\beta \omega_e/2)$ is the equilibrium polarization, $\Delta_0 = \omega_e - \omw$ and 
we assumed that the microwaves only act on the resonating packet.
Note that, for conserved quantities of \eqref{hamiltoniantoy}, as the energy and the total magnetization,
the balance of the flows has a simple form since it reduces to the exchanges with the bath and microwaves.

These results  retrace the traditional prediction  obtained within 
the phenomenological Ansatz of Eq.~\eqref{ansatz} proposed by Borghini \cite{borghini1968spin}. 
Here, Eq.~\eqref{ansatz} naturally emerges once the strong suppression of fluctuations, characteristic of 
the ETH phase, has been assumed. 
However, the qualitative approach to hyperpolarization provided by this toy 
model largely underestimate  the spin temperature value and 
hides its dependence from the microscopical parameters ($U, T_1,\ldots$) 
\cite{C3CP44667K}.

\begin{figure}[ht]
\includegraphics[width=1.0\columnwidth]{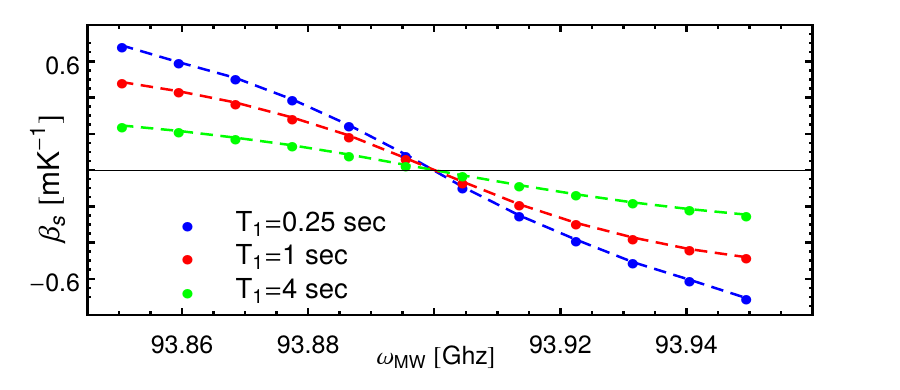}
\caption{\label{gadolinium}
$T_1$ shortening effect. The spin temperature $\beta_s$,
obtained from the fit of the electron polarizations (see Fig.~\ref{polarization}),
is shown versus $\omw$ with $U=15$ Mhz and different values of the relaxation time $T_1$: 
the spin temperature is smaller when relaxation is faster, consistently with the 
observed increase in the hyperpolarization efficiency \cite{ardenkjaer2008dynamic, 2012JPCA.116.5129L}.
}
\end{figure}

 A richer scenario emerges instead from the exact diagonalization of 
Eq.~\eqref{hamiltonian}, where
 a stronger hyperpolarization enhancement is observed approaching the MBL 
transition (see Fig.~\ref{polarization})
 and can be even amplified decreasing the relaxation time $T_1$ (see 
Fig.~\ref{gadolinium}).
 Both effects agree with two well-known experiments, which fall beyond the 
applicability of
 the Borghini model \cite{C3CP52534A}. The first showed that the enhancement 
occurs 
 only at relatively low radical concentrations \cite{johannesson2009dynamic}, 
 and therefore at weaker dipolar interactions. In the second,
 the addition of gadolinium complexes was used to induce a reduction 
of the relaxation time $T_1$ \cite{ardenkjaer2008dynamic, 2012JPCA.116.5129L}. This, in 
turn, improved the signal enhancement 
 and gadolinium is now commonly exploited in standard
 protocols for DNP sample preparation.

 \paragraph{Concluding remarks. ---}
 We presented a simple model for the study of DNP, providing
a realistic dependence on  the tunable parameters. The concept of out-of-equilibrium spin-temperature
emerges naturally as a macroscopic manifestation of the ETH for the electron spin hamiltonian.

 Our study candidates DNP as a good ground 
 for the direct observation of the MBL transition and its dynamical phase 
diagram.
 Two key advantages play in favor of this experimental setting. 
 The first is that the two relevant control parameters for the transition are 
tunable: 
  $U$ depends on the radical concentration and $\Delta \omega_e$ is 
 proportional to the external magnetic field. 
 The second is that the system does not require being isolated during the 
characteristic observation time,
 but, rather, that the relaxation  is sufficiently slow to allow the pure 
quantum behavior to settle. 
 Note that, in the past, the spin-temperature polarization profile  was
 already  experimentally observed in \cite{atsarkin1970verification} for 
increasingly  g-factors anisotropy, up to a critical value where
 the hole-burning profile popped out. 
The possibility of performing experiments precisely aimed at
the observation of the elusive critical regime of 
the MBL is therefore concrete and promising.
Moreover, the tunability of external parameters
may allow the exploration of the phase diagram,
even in regimes where the physics of spin glasses becomes relevant \cite{bapst2013quantum, laumann2014many}.


\nocite{mehta2004random}
\nocite{petruccione2002theory}
\nocite{martin1959theory}
\nocite{kubo1957statistical}


\begin{acknowledgments}

This work is supported by ``Investissements d'Avenir'' LabEx PALM 
(ANR-10-LABX-0039-PALM).
We thank M. Bauer, D. Bernard, P. Carretta,  S. Colombo Serra, M. Filibian, Y. 
Hovav, L. Mazza and F. Tedoldi for useful discussions. I. Rodriguez Arias and C. Zankoc are acknowledged for
the careful reading of the manuscript.

\end{acknowledgments}




\bibliography{dnp}
\newpage


\section*{Supplementary material}

# Supporting information
# Dynamic nuclear polarization and the paradox of Quantum Thermalization

In the supporting information we provide the technical details of our computations. In Sec. I we write the master equation for the occupation probabilities of the exact eigenstates. In Sec. II we compute explicitly the polarization profile in absence of dipolar interactions, where $P_e(\omega)$ displays an hole burning shape. In Sec. III, we derive the simplified master equation for the occupation probabilities of the sectors of the hamiltonian $\hat{H}_{\text{toy}}$ defined in the main text. Finally, in Sec. IV we specify the numerical procedures and the parameters used to produce the figures of the main text.

The form of the transition rates, used in our calculations, is analogous to the one introduced by Hovav and collaborators, in Ref. [1]. In order to keep the paper self-contained, we derive, in the appendix, the relaxation scheme within the Lindblad formalism. In particular, the analysis of the energy exchange with the thermal bath is presented in App. A; the processes induced by the microwave irradiation are treated, within the rotating wave approximation, in App. B. Finally in App. C we show how to reduce the time evolution of the density matrix to a master equation for the occupation probabilities.

## I. MASTER EQUATION AND POLARIZATION PROFILE

The spin dynamics is governed by a master equation for occupation probabilities, $p_n$, of the $2^N$ exact eigenstates of $\hat{H}$

$$\frac{dp_n}{dt} = \sum_{n' \neq n} \left[ h(-\omega_{nn'}) W_{n,n'}^{\text{bath}} + W_{n,n'}^{\text{MW}} \right] p_{n'} - \left[ h(\omega_{nn'}) W_{n,n'}^{\text{bath}} + W_{n,n'}^{\text{MW}} \right] p_n \tag{1}$$

where the rate of microwave induced process writes

$$W_{n,n'}^{\text{MW}} = \frac{4\omega_1^2 T_2}{1 + (T_2 \Delta\omega_{nn'})^2} |\langle n| \hat{S}_x |n'\rangle|^2, \quad \Delta\omega_{n,n'} = \epsilon_n - \epsilon_{n'} - (s_z^n - s_z^{n'})\omega_{\text{MW}}, \tag{2}$$

with $\epsilon_n, \epsilon_{n'}$ and $s_z^n, s_z^{n'}$ eigenvalues of the operators $\hat{H}$ and $\hat{S}_z$ on $|n\rangle, |n'\rangle$ and $\hat{S}_x = \sum_{j=1}^{N} \hat{S}_x^j$ (see App. B and App. C). In presence of a strong magnetic field: $\Delta\omega_{n,n'}^2 = (|\epsilon_n - \epsilon_{n'}| - \omega_{MW})^2$, as reported in the main text. The transitions with the thermal bath are governed by the function $h(\omega) = \frac{e^{\beta\omega}}{e^{\beta\omega}+1}$, which assures the detailed balance condition, and by the rates (see App. A)

$$W_{n,n'}^{\text{bath}} = \frac{2}{T_1} \sum_{\substack{j=1 \\ \alpha=x,y,z}}^{N} |\langle n| \hat{S}_\alpha^j |n'\rangle|^2. \tag{3}$$

The local polarization $P_e(\omega, t)$ of the electron spin under a magnetic field $\omega = \omega_e + \Delta_j$ writes as

$$P_e(\omega, t) = 2 \text{Tr}[\rho(t) \hat{S}_z^j] \simeq 2 \sum_n p_n(t) \langle n| \hat{S}_z^j |n\rangle. \tag{4}$$

where we approximated the density matrix $\rho(t)$ with its diagonal.

## II. BLOCH EQUATIONS AND HOLE BURNING SHAPE

In absence of dipolar interactions, the eigenstates of the spin system are products of local eigenstates of $\hat{S}_z^j$,

$$|a\rangle = |\uparrow\downarrow\downarrow \ldots\rangle, \tag{5}$$

and the transition rates connect states that differ for a single spin flip, say the spin $j$:

$$W_{n\to n'} = W_{\uparrow\to\downarrow}^j = \frac{h(\omega_e + \Delta_j)}{T_1} + \frac{\omega_1^2 T_2}{\Delta\omega^2 T_2^2 + 1} \tag{6}$$

where $\Delta\omega = \omega_e + \Delta_j - \omega_{\text{MW}}$. In this limit the probabilities $p_n$ factorize in the single spin probabilities, thus, the master equation (1) simplifies

$$\frac{dp_{j=\uparrow}}{dt} = W^j_{\downarrow\to\uparrow} p_{j=\downarrow}(t) - W^j_{\uparrow\to\downarrow} p_{j=\uparrow}(t)$$

$$\frac{dp_{j=\downarrow}}{dt} = W^j_{\uparrow\to\downarrow} p_{j=\uparrow}(t) - W^j_{\downarrow\to\uparrow} p_{j=\downarrow}(t)$$

and can be re-written as the Bloch equation for the local polarization $P_e(\omega, t) = p_{j=\uparrow}(t) - p_{j=\downarrow}(t)$ at frequency $\omega = \omega_e + \Delta_j$:

$$\frac{dP_e(\omega,t)}{dt} = -\frac{2\omega_1^2 T_2 P_e(\omega,t)}{\Delta\omega^2 T_2^2 + 1} + \frac{P_e^0(\omega) - P_e(\omega,t)}{T_1} \tag{7}$$

where $P_e^0(\omega) = -\tanh(\beta\omega/2)$. The stationary solution corresponds to the hole burning shape of the irradiated EPR spectrum:

$$P_e(\omega) = \frac{(1 + \Delta\omega^2 T_2^2) P_e^0(\omega)}{1 + \Delta\omega^2 T_2^2 + 2T_1 T_2 \omega_1^2} \tag{8}$$

with $\Delta\omega = \omega - \omega_{\text{MW}}$.

### III. MASTER EQUATION FOR THE TOY MODEL

We now specialize the master equation (1) to the toy model defined by Eq. (4) of the main text. The spectrum of $\hat{H}_{\text{toy}}$, for $\eta = 0$, is highly degenerate and since $[\hat{H}_{\text{toy}}, \hat{S}_z] = 0$, we label as $V_{\epsilon, s_z}$ the common eigenspace of $\hat{H}_{\text{toy}}$ and $\hat{S}_z$ with eigenvalues, respectively, $\epsilon$ and $s_z$. When a small $\eta > 0$ is turned on, the matrix $\hat{H}_{\text{int}}$ resolves the degeneration and selects a non-degenerate basis in each subspace, so that for any $|n\rangle \in V \equiv V_{\epsilon, s_z}$

$$|n\rangle = \sum_{a=1}^d c_a |a\rangle \tag{9}$$

where the states $|a\rangle$ are product of local eigenstates of $\hat{S}_z^i$, as in (5). The sum runs over the states inside the subspace $V$, and $d = \dim(V)$. Choosing $\hat{H}_{\text{int}}$ from the gaussian orthogonal ensemble (GOE), the coefficients $c_a$ are distributed according to the Porter-Thomas distribution [2]. In other words, for each eigenstate in a subspace $V$, they are chosen indipendently from the standard normal distribution, with the additional constraint of normalization

$$\sum_{\phi=1}^d (c_a)^2 = 1 \tag{10}$$

so that $\overline{c_a^2} = 1/d$, where the average is over the random matrix ensemble.

For $\eta$ small enough, the energies of different sectors are well separated so that $\epsilon$ remains a good quantum number. In this limit the master equation (1) connects only states belonging to subspaces differing by a single spin-flip. If we consider, for example, a spin $i$ in the packet $p$ that flips from up to down, we have a transition between a subspace $V \equiv V_{\epsilon, s_z}$ and a subspace $V' = V_{\epsilon - \hbar\omega_p, s_z - 1}$, with $\omega_p = \omega_e + \Delta_p$. Taking $|n\rangle, |n'\rangle$ in $V, V'$, the matrix element associated to this transition writes

$$\langle n| \hat{S}_x^i |n'\rangle = \sum_{aa'} c_a c'_{a'} \langle a| \hat{S}_x^i |a'\rangle = \frac{1}{2} \sum_{a \in V_{i=\uparrow}} c_a c'_{\tilde{a}} \tag{11}$$

where we decomposed $V = V_{i=\uparrow} \cup V_{i=\downarrow}$ according to the value of the spin $i$ and $|\tilde{a}\rangle$ is obtained from $|a\rangle$ by flipping $i$. From (11), we see that the matrix element converges to a gaussian number with zero average and variance

$$\overline{|\langle n| \hat{S}_x^i |n'\rangle|^2} = \frac{\dim V_{i=\uparrow}}{4dd'} = \frac{\mathcal{P}_{p,\uparrow}}{4d'} \tag{12}$$

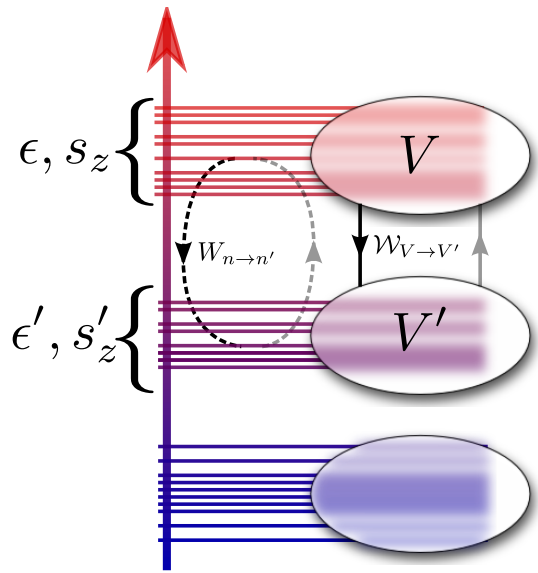

FIG. 1. Sketch of the coarse-grained rate equation for the toy model defined in Eq. (4) of the main text. The rate of the transitions between two ergodic eigenstates $|n\rangle, |n'\rangle$ depends only on the two subspaces $V, V'$. The time-evolution of the system is then described by a master equation between subspaces.

where $d' = \dim(V')$ and $\mathcal{P}_{p,\uparrow}$ is the probability that, for any state in $V$, a spin in the packet $p$ is up. Neglecting the small fluctuations between states, we obtain that the rate of the transitions $W_{n \to n'}$ depends only on the two subspaces that contain $n$ and $n'$, in this case $W_{n \to n'} \longrightarrow W_{V \to V'}$. The master equation (1) between states can then be reduced to a coarse-grained dynamics between the occupation probabilities of each subspace, $p_V = \sum_{n \in V} p_n$. The coarse-grained rates between subspaces write as (see Fig. 1)

$$\mathcal{W}_{V \to V'} = d' W_{V \to V'} = N_p \mathcal{P}_{p,\uparrow} \left[ \frac{h(-\hbar \omega_p)}{T_1} + \frac{2\omega_1^2 T_2}{1 + T_2^2(\omega_p - \omega_{\text{MW}})^2} \right] \tag{13}$$

Note that, the validity of Eq. (13) strongly relies on the ergodicity of the eigenstates. For instance, if the eigenstates were completely factorized, the rate of the transitions between $|a\rangle$ and $|a'\rangle$ would retain finite fluctuations since the matrix elements $|\langle a| \hat{S}_x^i |a'\rangle|^2$ are either zero or one.

Although the master equation described by Eq. (13) cannot be solved analytically, it allows the study of much larger systems as the number of subspaces grows only polynomially with $N$. Moreover, imposing that the $d\langle \hat{H} \rangle/dt = 0$ and $d\langle \hat{S}_z \rangle/dt = 0$, we obtain Eqs.(6) in the text.

## IV. METHODS

The model defined by Eqs. (2,3) of the main text was solved with the following parameters

| $T_1$ (sec.) | $T_2$ (sec.) | $\omega_1$ ($2\pi$ Ghz) | $\beta$ (K$^{-1}$) | $U$ (Mhz) | $\omega_e$ (Ghz) | $\Delta\omega_e$ (Mhz) |
|---|---|---|---|---|---|---|
| $0.25 \div 4.$ | $10^{-6}$ | $0.25 \times 10^{-4}$ | 0.833 | $2.0 \div 45.0$ | 93.9 | 108.0 |

and the local magnetic fields were chosen as

$$\Delta_i = \Delta\omega_e \left( \frac{2i - N - 1}{2N} \right) \tag{14}$$

where $N$ is the number of electrons. This choice corresponds in the limit of large $N$ to a uniform distribution $f(\omega) \simeq 1/\Delta\omega_e$. In order to obtain the polarization profile employed in Fig. 2, we performed the following steps:

- Numerical diagonalization of $\hat{H}$ in each block at fixed total magnetization.
- Solution of the linear system determining the stationary $p_n$ from the master equation (1).

- Evaluation of the stationary polarizations, $P_e(\omega)$, using (4).

The resulting polarization for each $\omega$ are then averaged over at least 50 realizations.

For Fig. 3 (left), the distribution $P$ of diagonal elements $\langle n| \hat{S}_z^i |n\rangle$ was computed sampling, across $i$ and $10^3$ disorder realisations, all the eigenstates in the sector of vanishing total polarization. Then, the coloured regions were obtained, selecting for each energy window $(\epsilon, \epsilon + \Delta\epsilon)$, the area with maximal concentration containing half probability. For Fig. 3 (middle), we focused on the sector of zero total magnetization, i.e. $s_z = 0$, of $\hat{H}$ and computed the variation of the local polarization between pairs of adjacent eigenstates, defined as:

$$\delta M_n^i \equiv \langle n+1| S_z^i |n+1\rangle - \langle n| S_z^i |n\rangle . \tag{15}$$

The expectation value $\overline{|\delta M_n^i|}$ was obtained by averaging over the spin index $i$, $n$ belonging to the middle third of the spectrum, and a number of realizations ranging from $10^2$ to $10^4$. For Fig. 3 (right), we determined numerically the stationary state of the coarse-grained master equation, defined by the rates in (13). We considered 7 packets resonating at frequencies $\omega_p = \omega_e + \Delta\omega_e(2p-8)/14$ with $N_p = 4, 8, 12, 16, 12, 8, 4$. The probabilities $\mathcal{P}_{p,\uparrow}$ in (13) were computed by exact enumeration of the configurations of the subspaces.

### Appendix A: Lindblad equation in weak coupling approximation

We summarize here the technique we use to treat the interaction of the spin system with the lattice bath. The full Hamiltonian is composed by three terms

$$\hat{\mathcal{H}} = \hat{H}_\text{S} + \hat{H}_\text{B} + \lambda \hat{H}_\text{S-B} \tag{A1}$$

where $\hat{H}_\text{S}$ is the spin system Hamiltonian, as given in Eq. (2) of the main text, $\hat{H}_\text{B}$ controls the evolution of the bath and $\hat{H}_\text{S-B}$ describes the coupling of the system with the bath; the parameter $\lambda$ measures the strength of this coupling. The most relevant contributions to $\hat{H}_\text{S-B}$ involve single-spin operators

$$\hat{H}_\text{S-B} = 2 \sum_{\substack{j=1 \\ \alpha=x,y,z}}^{N} \hat{S}_\alpha^j \hat{\Phi}_\alpha^j \tag{A2}$$

where the operators $\hat{\Phi}_\alpha^j$ describe the excitations associated with the slow molecular modes localized around the electron spin $j$. We are interested in the effective long-time dynamics of the spin system when $\lambda$ is small, i.e. the interactions with the lattice are negligible with respect to the energy scales of the spin system. Within the Born-Markov approximation [3] the role of the bath is encoded in the autocorrelation function $\langle \hat{\Phi}_\alpha^k(t_0) \hat{\Phi}_\beta^j(t_0 + t)\rangle_\text{bath}$, which being stationary depends only on $t$. For simplicity, we assume locality, homogeneity and isotropy so that we can restrict to $\alpha = \beta$, $j = k$ and drop indexes. Then, we define the spectral density

$$J(\omega) = \int_{-\infty}^{\infty} e^{i\omega t} \text{Tr}[\rho_\text{B} \hat{\Phi} e^{-i\hat{H}_\text{B}t} \hat{\Phi} e^{i\hat{H}_\text{B}t}] , \tag{A3}$$

where $\rho_\text{B}$ is the density matrix of the bath. The quantum evolution for the density matrix of the spin system, $\rho \equiv \rho_\text{S}$, takes the form [4]

$$\frac{d\rho}{dt} = -i[\hat{H}_\text{S}, \rho] + \mathcal{L}\rho = -i[\hat{H}_\text{S}, \rho] + \frac{\lambda^2}{2} \sum_{\substack{j=1 \\ \alpha=x,y,z}}^{N} \sum_\omega J(\omega) \left\{ [\hat{S}_\alpha^j(\omega)\rho, \hat{S}_\alpha^j(-\omega)] + [\hat{S}_\alpha^j(\omega), \rho\hat{S}_\alpha^j(-\omega)] \right\} , \tag{A4}$$

where $\mathcal{L}$ is the Lindbladian, which preserves positivity and trace of the density matrix. The sum over $\omega$ runs over all the energy differences of the exact eigenstates of $\hat{H}_\text{S}$. The presence of random dipolar coupling in $\hat{H}_\text{S}$ ensures non-degenerate gaps so that

$$\hat{S}_\alpha^j(\omega_{nn'}) = |n'\rangle \langle n'| \hat{S}_\alpha^j |n\rangle \langle n|, \quad \omega_{nn'} = \epsilon_n - \epsilon_{n'}, \tag{A5}$$

with $\epsilon_n$, $\epsilon_{n'}$ the energy levels associated to the eigenstates $|n\rangle, |n'\rangle$. Let us observe that in the weak coupling limit, the bath influences the dynamics by allowing transitions between the exact eigenstates of the system Hamiltonian $\hat{H}_\text{S}$.

Since the bath is at equilibrium, it is easy to check, by replacing $t \to t - i\beta$ in (A3), the KMS condition [5, 6]

$$J(\omega) = e^{\beta \omega} J(-\omega) \,. \tag{A6}$$

This condition implies the detailed balance of the rates and therefore the convergence toward equilibrium $\rho = e^{-\beta \hat{H}_S}/Z$ where $\beta^{-1}$ is the bath temperature.

It is convenient to rewrite Eq. (A4) separating the evolution of diagonal and off-diagonal terms of $\rho$. For the diagonal terms, we have

$$\frac{d\rho_{nn}}{dt} = \sum_{n'} [h(-\omega_{nn'})\rho_{n'n'} - h(\omega_{nn'})\rho_{nn}] \, W_{n,n'}^{\text{bath}} \tag{A7}$$

where $h(\omega) = \frac{e^{\beta\omega}}{e^{\beta\omega}+1}$, satisfying (A6), and the rate of the transitions between the levels $n$ and $n'$ is given by

$$W_{n,n'}^{\text{bath}} = \frac{2}{T_1} \sum_{\substack{j=1 \\ \alpha=x,y,z}}^{N} |\langle n| \hat{S}_\alpha^j |n'\rangle|^2. \tag{A8}$$

The relaxation time, $T_1$, fixes the time scale of the electronic single-spin transitions. Its value depends on $J(\omega)$ and the coupling strength $\lambda$. The transitions involving $\hat{S}_{x,y}$ correspond to a change of the total polarization and their time scale can be estimated by the experimental value of the electron relaxation time, $T_1 \sim 1$ sec. On the contrary a direct measure of the transitions involving $\hat{S}_z$ is difficult and they are usually neglected [1]. Here we assume that their characteristic time scale remains of the order of 1 sec. For the off diagonal terms we have

$$\frac{d\rho_{nm}}{dt} = -i(\epsilon_n - \epsilon_m)\rho_{nm} - \frac{\rho_{nm}}{T_{2,nm}}. \tag{A9}$$

The exact expression of $T_{2,nm}$ in terms of the spectral function can be derived from Eq. (A4). However since the off diagonal decay rate is generically much faster then the relaxation time $T_1$ we assume for simplicity $T_{2,nm} = T_2$, with the experimental decoherence time $T_2 = 10^{-6}$ sec. [1].

### Appendix B: Microwave irradiation in rotating wave approximation

We irradiate the system with a microwave field

$$\hat{H}_{\text{MW}} = 2\omega_1 \cos(\omega_{\text{MW}} t)\hat{S}_x \tag{B1}$$

where $\omega_1$ is the intensity of the field, $\omega_{\text{MW}}$ its frequency of the order of 100 Ghz and $\hat{S}_x = \sum_{j=1}^{N} \hat{S}_x^j$. The evolution equation (A4) is modified as

$$\frac{d\rho}{dt} = -i[\hat{H}_S, \rho] - i[\hat{H}_{\text{MW}}, \rho] + \mathcal{L}\rho. \tag{B2}$$

Since the microwave power is much smaller than the energy scales of the spin system, the Lindbladian, $\mathcal{L}$, is not affected by the microwave field. Eq. (B2) is now time-dependent and it is convenient to employ the so-called *rotating wave approximation* (RWA). We define the density matrix in the rotating frame as $\rho^{(r)} = \hat{U}(t)\rho\,\hat{U}^\dagger(t)$, with $\hat{U}(t) = \exp(i\hat{S}_z \omega_{\text{MW}} t)$ and $\hat{S}_z = \sum_j \hat{S}_z^j$. In this way, since $[\hat{H}_S, \hat{S}_z] = 0$, Eq. (B2) writes

$$\frac{d\rho^{(r)}}{dt} = -i[\hat{H}_S^{(r)}, \rho^{(r)}] - i[\hat{U}(t)\hat{H}_{\text{MW}}\hat{U}^\dagger(t), \rho^{(r)}] + \hat{U}(t)\left(\mathcal{L}\rho\right)\hat{U}^\dagger(t) \tag{B3}$$

where $H^{(r)} = \hat{H}_S - \omega_{\text{MW}}\hat{S}_z$. Moreover we have

$$\hat{H}_{\text{MW}}^{(r)} = \hat{U}(t)\hat{H}_{\text{MW}}\hat{U}^\dagger(t) = \omega_1 \hat{S}_x + \sum_{j=1}^{N} e^{2i\hat{S}_z^j \omega_{\text{MW}} t} \omega_1 \hat{S}_x^j e^{-2i\hat{S}_z^j \omega_{\text{MW}} t} \simeq \omega_1 \hat{S}_x$$

where the fast oscillating term has been neglected. Since the total magnetizanion $\hat{S}_z$ is a good quantum number, we have from (A5)

$$\hat{U}(t)\hat{S}_\alpha^j(\pm\omega)\hat{U}^\dagger(t) = e^{\pm i(s_z^n - s_z^{n'})\omega_{\text{MW}} t}\hat{S}_\alpha^j(\pm\omega) \tag{B4}$$

where $s_z^n, s_z^{n'}$ are the eigenvalues of $\hat{S}_z$ on $|n\rangle, |n'\rangle$. It follows that the last term of Eq.(B3) can be rewritten as $\mathcal{L}\rho^{(r)}$. and in conclusion, adopting the RWA approximation, Eq. (B2) becomes

$$\frac{d\rho^{(r)}}{dt} = -i[H^{(r)}, \rho^{(r)}] - i\omega_1[\hat{S}_x, \rho^{(r)}] + \mathcal{L}\rho^{(r)} \ . \tag{B5}$$

Separating the diagonal and off-diagonal dynamics, it can be rewritten as

$$\frac{d\rho_{nn}^{(r)}}{dt} = -i\omega_1 \sum_{n'} \langle n|S_x|n'\rangle (\rho_{n'n}^{(r)} - \rho_{nn'}^{(r)}) + \sum_{n'} \left[h(-\omega_{nn'})\rho_{n'n'}^{(r)} - h(\omega_{nn'})\rho_{nn}^{(r)}\right] W_{n,n'}^{\text{bath}} \tag{B6}$$

$$\frac{d\rho_{nm}^{(r)}}{dt} = -\left(i\Delta\omega_{nm} + \frac{1}{T_2}\right)\rho_{nm}^{(r)} - i\omega_1 \sum_{n'} \langle n|\hat{S}_x|n'\rangle \rho_{n'm}^{(r)} - \langle n'|S_x|m\rangle \rho_{nn'}^{(r)}$$

where $\Delta\omega_{nm} = \epsilon_n - \epsilon_m - (s_z^n - s_z^m)\omega_{\text{MW}}$.

## Appendix C: Master equation in the fast dephasing limit

In our system, the dephasing time $T_2$ is much faster than the relaxation time $T_1$ ($T_2/T_1 \simeq 10^{-6}$) and the quantum evolution can be reduced to a master equation for the diagonal elements of the density matrix in the basis which diagonalises $\hat{H}_S$. We start rewriting (B6) as

$$\frac{d\rho^{(r)}}{dt} = \mathbf{L}_0 \rho^{(r)} + \mathbf{L}_1 \rho^{(r)} \tag{C1}$$

where $\mathbf{L}_{0,1}$ are superoperators acting linearly on the space of densitiy matrices. $\mathbf{L}_0$ is the dominant part involving contributions from $T_2^{-1}$ and $\Delta\omega_{nn'}$, while $\mathbf{L}_1$ contains the remaining contributions. We denote as $e_{nn'}$ the matrices with a single one in position $n, n'$. They correspond to the right eigenvectors of $\mathbf{L}_0$ with eigenvalues:

$$\mathbf{L}_0 e_{nn'} = \begin{cases} -(T_2^{-1} + i\Delta\omega_{nn'}) & n \neq n', \\ 0 & n = n'. \end{cases} \tag{C2}$$

In the long-time limit the dynamics is restricted to the subspace with eigenvalue 0, corresponding to the diagonal density matrices. The effective evolution in this subspace due to the presence of $\mathbf{L}_1$ can be derived by standard perturbation theory. By setting $p_n = \rho_{nn}^{(r)} = \rho_{nn}$ and indicating as $(\mathbf{L}_1)_{nn',mm'}$ the components of the superoperator $\mathbf{L}_1$ on the basis $e_{nn'}$, we obtain

$$\frac{dp_n}{dt} = \sum_{n'} \left[(\mathbf{L}_1)_{nn,n'n'} + \sum_{m \neq m'} \frac{(\mathbf{L}_1)_{nn,mm'}(\mathbf{L}_1)_{mm',n'n'}}{T_{2,mm'}^{-1} + i\Delta\omega_{mm'}}\right] p_{n'} \tag{C3}$$

After some algebra we get the final expression given in Eq.(1)

---



\end{document}